\documentstyle[12pt]{article}
\newlength{\extraspace}
\newlength{\extraspaces}
\newcommand{\be}{\begin{equation}
\addtolength{\abovedisplayskip}{\extraspaces}
\addtolength{\belowdisplayskip}{\extraspaces}
\addtolength{\abovedisplayshortskip}{\extraspace}
\addtolength{\belowdisplayshortskip}{\extraspace}}
\newcommand{\ee}{\end{equation}}
\newcommand{\ba}{\begin{eqnarray}
\addtolength{\abovedisplayskip}{\extraspaces}
\addtolength{\belowdisplayskip}{\extraspaces}
\addtolength{\abovedisplayshortskip}{\extraspace}
\addtolength{\belowdisplayshortskip}{\extraspace}}
\newcommand{\ea}{\end{eqnarray}}
\newcommand{\ban}{\setcounter{enumi}{\value{equation}} 
\addtocounter{enumi}{1}
\setcounter{equation}{0}
\renewcommand{\theequation}{\theenumi\alph{equation}}
\begin{eqnarray}
\addtolength{\abovedisplayskip}{\extraspaces}
\addtolength{\belowdisplayskip}{\extraspaces}
\addtolength{\abovedisplayshortskip}{\extraspace}
\addtolength{\belowdisplayshortskip}{\extraspace}}
\newcommand{\ean}{\end{eqnarray}
\setcounter{equation}{\value{enumi}}
\renewcommand{\theequation}{\arabic{equation}}}
\newcommand{\nonu}{\nonumber \\[.5mm]}
\newcommand{\an}{&\!\!\!}
\newcommand{\dl}{\delta{\cal L}}
\newcommand{\lt}{{\cal L}_t}
\newcommand{\E}{\tilde{E}{}}
\newcommand{\A}{A^{\prime}{}}
\newcommand{\de}{\delta}
\newcommand{\ep}{\epsilon}
\newcommand{\ga}{\gamma}

\newcommand{\La}{\Lambda}
\newcommand{\bp}{\bar{\psi}}
\newcommand{\M}{M {\!\!\!\!\!_{_{ \sim}}}\,}
\newcommand{\N}{N {\!\!\!\!\!_{_{ \sim}}}\,}
\newcommand{\pa}{\partial}
\newcommand{\fr}{\frac{1}{2}}
\newcommand{\n}{(n\cdot\ga)}
\newcommand{\ap}{\approx}
\newcommand{\da}{\dagger}
\newcommand{\Ha}{H^{\prime}{}}
\newcommand{\G}{G^{\prime}{}}
\newcommand{\D}{D^{\prime}{}}
\newcommand{\F}{F^{\prime}{}}
\newcommand{\im}{{\rm{Im}}}
\begin{document}
\addtolength{\baselineskip}{.7mm}
\begin{flushright}
STUPP-96-148 \\ September, 1996
\end{flushright}
\vspace{.6cm}
\begin{center}
{\large{\bf{Reality conditions for (2+1)-dimensional gravity 
               coupled with the Dirac field}}} \\[20mm]
{\sc Manabu Sawaguchi\footnote{e-mail address: sawa@th.phy.saitama-u.ac.jp} 
and Chol-Bu Kim\footnote{e-mail address: cbkim@th.phy.saitama-u.ac.jp}} 
\\[12mm]
{\it Physics Department, Saitama University \\[2mm]
Urawa, Saitama 338, Japan} \\[20mm]
{\bf Abstract}\\[10mm]
{\parbox{13cm}{\hspace{5mm}
The canonical formalism of three dimensional gravity 
coupled with the Dirac field is considered.
We introduce complex variables 
to simplify the Dirac brackets of canonical variables 
and examine the canonical structure of the theory.
We discuss the reality conditions which guarantee 
the equivalence between the complex and real theory.
}} 
\end{center}
\vfill

\newpage
\section{Introduction}

In the canonical formalism of general relativity, 
the spatial metric and its canonical momentum have been considered 
as canonical variables. However, since constraints are non-polynomial form, 
it is difficult to solve the constraint equations. 
To avoid this difficulty, Ashtekar has introduced new variables which 
consist of a complexified connection and a densitized triad 
as canonical variables, and has shown that all constraints become of a 
polynomial form~\cite{A}. 
Inclusion of matter fields has been treated in the Ashtekar formalism, 
and consequences similar to those for pure gravity were obtained~\cite{ART}. 
In Ashtekar formalism, to recover real general relativity, 
the reality conditions must be imposed on the canonical variables~\cite{I,YS}. 

In 2+1 dimensions, the canonical formalism which is an extension 
of Ashtekar formalism has been proposed 
in the pure gravity~\cite{AHRSS} and the $N=2$ supergravity~\cite{MN}. 
In the case of including the Dirac field, 
Kim {\it et al}~\cite{KSY} have shown 
that the constraint structure is similar to the 
3+1 dimensional gravity~\cite{ART}, and have found a new physical observable 
and its eigenstate. 
In~\cite{KY}, the solution of all quantum constraints 
based on the loop representation has been found. 
In this theory, 
since the spinor action consists of derivative of $\psi$ only, 
the total action becomes non-Hermitian. 
Vergeles~\cite{Ver} has also carried out the quantization 
on the basis of the dynamic quantization method, 
in which the spinor action contains derivatives of both $\psi$ and $\bp$. 
Thus in contrast with the non-Hermitian case~\cite{KSY}, 
the real total action is used. 
In this formalism, 
the Dirac field modes with gauge invariant creation and 
annihilation operators were selected, 
and the gauge invariant states were constructed by 
using the gauge invariant fermion creation operators similarly to 
the usual construction of states in any Fock space.

In this paper we consider 2+1 dimensional gravity coupled with the 
Dirac field. We start with a real action. 
First, we carry out a Hamiltonian formulation. 
By solving the second class constraints, we obtain 
the Dirac brackets of canonical variables, which are not a simple form. 
We also present the algebra of the constraints, in which 
some quadratic terms of the Gauss-law constraint appear. 
Next, we introduce new variables which obey simpler Dirac brackets. 
We show that the canonical structure of the theory is similar to 
the case of the non-Hermitian action in 2+1 dimensions~\cite{KSY}, 
and also to the case of matter coupled gravity in 3+1 dimensions~\cite{ART}. 
Since the new variables are complex, the reality conditions must be 
imposed like in the 3+1 dimensional gravity. 
In section 4, we describe the reality conditions in the form of functionals 
of the canonical variables and their complex conjugate. 
It is shown that, 
by imposing the reality conditions, the correct 
number of degrees of freedom remains.
The last section is devoted to conclusions and comments. 

\section{Canonical formalism}

We start with the first-order gravity in 2+1 dimensions 
coupled with the Dirac field. The fundamental variables we use 
are the triad field $e_{\mu}^i$ and the dual spin connection
\be
A_{\mu}^i = -\fr \ep^{ijk} \omega_{\mu jk},
\ee
instead of the triad field and the usual spin connection $\omega_{\mu ij}$. 

The action is written as
\be
I = \int d^3x \left[
\frac{1}{2}\ep^{\mu\nu\rho}e^i_{\mu}F_{\nu\rho i}
- \frac{1}{2}e ( \bp \ga^{\mu} D_{\mu} \psi
               - D_{\mu} \bp \ga^{\mu} \psi )
- me\bp \psi
\right], 
\ee
where $F_{\nu\rho i} = \pa_{\nu}A_{\rho i} - \pa_{\rho}A_{\nu i}
- \ep_{ijk}A^j_{\nu}A^k_{\rho} $ 
is the curvature tensor of the spin connection, 
$\ep^{\mu\nu\rho}$ is the Levi-Civita antisymmetric tensor density and 
$D_{\mu} \psi = (\pa_{\mu} +\fr A_{\mu}^i\ga_i)\psi$. 
The matrices $\ga_i$ generate the SO(2,1) Lorentz group. 
The Dirac conjugation is defined by $\bp := i\psi^{\da}\ga_0$. 
We use Greek letters $\mu, \nu, \cdots$ for curved indices 
in three dimensions and 
Latin letters $i, j, \cdots$ for tangent space indices.
The spacetime metric has a signature $(-,+,+)$ and 
we use the convention $\ep^{012} = 1 = - \ep_{012}$.

We decompose the spacetime metric following the ADM formalism. 
We assume that the spacetime manifold $M$ has a topology 
$M = \Sigma \otimes R$, where $\Sigma$ is a compact 
two dimensional manifold. 
We choose a time coordinate $t$ on the manifold $M$ so that 
$M$ is foliated by two dimensional spacelike surfaces 
$\Sigma_t$ each with the topology of $\Sigma$. 
One can define a timelike unit vector $n^{\mu}$ with 
$n^{\mu}n^{\nu} g_{\mu\nu} = -1$ which is normal 
to the $\Sigma_t$ and a smooth time vector field $t^{\mu}$ 
which is chosen such that $t^{\mu} \nabla_{\mu} t = 1$. 
We then define the spatial metric $q_{\mu\nu}$ by 
$q_{\mu\nu} = g_{\mu\nu} + n_{\mu}n_{\nu}; 
n^{\mu}q_{\mu\nu} = 0$, 
the timelike part $n^i$of $e^i_{\mu}$ by 
$n^i \equiv n^{\mu} e^i_{\mu}$ 
and the projected part of $e^i_{\mu}$ into 
$\Sigma_t$ by 
$E^i_{\mu} \equiv e^i_{\nu} (g^{\nu}_{\mu}
+ n_{\mu}n^{\nu}); n^{\mu}E^i_{\mu} = 0, 
q_{\mu\nu} = E^i_{\mu}E^j_{\nu} \eta_{ij}$. 
The Levi-Civita density $\ep^{\mu\nu\rho}$ 
is related to a density on $\Sigma_t, \ep^{\mu\nu}$ by 
$\ep^{\mu\nu\rho} = 3Nn^{[\mu}\ep^{\nu\rho]}$.
The decomposed action takes the form 
\ba
I = \int d^3x
[
\an\E^{\mu i}\an\!\!(\lt A_{\mu i}) + \fr E\bp \n (\lt \psi)
- \fr E(\lt \bp) \n \psi  \nonu
\an+\an \!\N \{ \fr \ep^{ijk}\E^{\mu}_i\E^{\nu}_j F_{\mu\nu k}
             + \fr E\ep^{ijk}\E^{\mu}_in_j\Psi_{\mu k} 
             - mE^2\bp\psi \} \nonu
\an+\an \!N^{\mu} \{ -\E^{\nu i} F_{\mu\nu i} 
                - \fr En_i\Psi^i_{\mu} \} \nonu
\an+\an \!A^i_0 \{ D_{\mu}\E^{\mu}_i + \fr En_i\bp\psi \}
]
\ea
where $\E^{\mu}_i$, $A^i_0$, $E$, $\N$, $\lt$ and $\Psi^i_{\mu}$ are 
a vector density on $\Sigma_t$ with 
$\E^{\mu}_i \equiv \ep^{\mu\nu}E_{\nu i}$, 
a time component of $A^i_{\mu}$ with $A^i_0 \equiv t^{\mu}A^i_{\mu}$, 
the determinant of $E_{\mu}^i$, 
a lapse with density weight $-1$, Lie derivative by $t^{\mu}$ and 
function defined as 
$\Psi^i_{\mu} := \bp\ga^i D_{\mu}\psi - (D_{\mu}\bp) \ga^i\psi$ 
respectively. 

For canonical treatment, we compute the canonical momenta. 
They are 
\ban
\Pi_{\mu}^i \an=\an \frac{\dl}{\de \dot{\E}{}^{\mu}_i} = 0, \ \ 
P^{\mu}_i = \frac{\dl}{\de \dot{A}{}_{\mu}^i} = \E^{\mu}_i \\
\pi \an=\an \frac{\dl}{\de \dot{\psi}} = -\fr E\bp\n, \ \ 
\bar{\pi} = \frac{\dl}{\de \dot{\bp}} = -\fr E\n\psi \\
\Pi_{\N} \an=\an \frac{\dl}{\de \dot{\N}} = 0, \ \ 
\Pi_{N\mu} = \frac{\dl}{\de \dot{N}{}^{\mu}} = 0, \ \ 
\Pi_{Ai} = \frac{\dl}{\de \dot{A}{}^i_0} = 0,\label{momenta}
\ean
where we use the convention that $\dot{q} \equiv \lt q$ and 
the canonical momenta of $\psi$ and $\bp$ are defined by the left derivative. 
Because these momenta do not depend on the velocities, 
these result in the primary constraints:
\ban
\Pi_{\mu}^i \an\ap\an 0, \ \ 
P^{\prime \mu}_i := P^{\mu}_i - \E^{\mu}_i \ap 0 \label{2nda} \\
\pi^{\prime} \an:=\an \pi + \fr E\bp\n \ap 0, \ \ 
\bar{\pi}^{\prime} := \bar{\pi} + \fr E\n\psi \ap 0 \label{2ndb} \\
\Pi_{\N} \an\ap\an 0, \ \ 
\Pi_{N\mu} \ap 0, \ \ 
\Pi_{Ai} \ap 0.\label{2ndc}
\ean
The consistency condition that the primary constraints are conserved 
requires secondary constraints,
\ban
H \an=\an \fr \ep^{ijk}\E^{\mu}_i\E^{\nu}_j F_{\mu\nu k}
             + \fr E\ep^{ijk}\E^{\mu}_in_j\Psi_{\mu k} 
             - mE^2\bp\psi \ap 0 \\
H_{\mu} \an=\an -\E^{\nu i} F_{\mu\nu i} 
                - \fr En_i \Psi^i_{\mu} \ap 0 \\
G_i \an=\an D_{\mu}\E^{\mu}_i + \fr En_i \bp\psi \ap 0 
\ean
and the velocity conditions which determine $\dot{\E}{}^{\mu}_i$, 
$\dot{A}{}_{\mu}^i$, $\dot{\psi}$ and $\dot{\bp}$.
The consistency conditions of the secondary constraints 
become the combination of the secondary constraints, 
from which it follows that there are no tertiary constraints. 
>From some algebra, it follows that the constraints (\ref{2nda}) and 
(\ref{2ndb}) are second class. The constraints (6) do not 
commute with constraints (5a,b). But when adding some linear combination of 
constraints (5a,b) to the constraints (6), 
we find that these are first class. 
The remaining constraints (\ref{2ndc}) tell us that the variables 
$\N$, $N^{\mu}$, $A^i_0$ and the corresponding momenta (\ref{momenta}) 
play a non-essential role in the Hamiltonian dynamics 
and become unphysical variables. 
Hereafter, we regard $\N$, $N^{\mu}$ and $A^i_0$ as Lagrange multipliers 
and the constraints (\ref{2ndc}) as zero strongly. 

Now we have to calculate the Dirac brackets to eliminate second 
class constraints. 
The non-vanishing brackets of the canonical variables are
\ba
\{\E^{\mu}_i(x),A_{\nu}^j(y)\} \an=\an q^{\mu}_{\nu}\eta^j_i \de^2(x,y) \nonu
\{\psi_{\alpha}(x),\bp_{\beta}(y)\} \an=\an 
     E^{-1}\n_{\alpha\beta}\de^2(x,y) \nonu
\{\psi_{\alpha}(x),A_{\mu}^i(y)\} \an=\an 
     \fr E^{-1}\ep_{\mu\nu}\ep^{ijk}\E^{\nu}_j(\n\ga_k\psi)_{\alpha} 
            \de^2(x,y)\nonu
\{\bp_{\alpha}(x),A_{\mu}^i(y)\} \an=\an 
     \fr E^{-1}\ep_{\mu\nu}\ep^{ijk}\E^{\nu}_j(\bp\ga_k\n)_{\alpha} 
            \de^2(x,y)\nonu
\{A^i_{\mu}(x),A^j_{\nu}(y)\} \an=\an \fr\ep_{\mu\nu}n^in^j\bp\psi
            \de^2(x,y)
\ea
Unfortunately, 
$A_{\mu}^i$ does not commute with $\psi$, $\bp$ and itself. 
In this respect the present theory differs 
from the case that starts with 
non-Hermitian action~\cite{KSY}, 
and the Dirac brackets are complicated at first sight.

After eliminating all the second class constraints, 
remaining first class constraints are 
\be
H \ap 0, \ H_{\mu} \ap 0, \ G_i \ap 0.\label{1st}
\ee
These constraints are called the Hamiltonian, the vector and 
the Gauss-law constraints respectively. 
The total Hamiltonian in the reduced phase space 
is described by these constraints as follows, 
\be
{\cal H} = -(\N H + N^{\mu}H_{\mu} + A^i_0 G_i).\label{Hamiltonian}
\ee
Note that, 
since constraints and multipliers contained in (\ref{Hamiltonian}) are real, 
this Hamiltonian is also real. 
So we do not have to consider the reality conditions. 

Now we discuss the algebra of the constraints.
Instead of the constraints (\ref{1st}), 
we use the following constraints smeared with suitable 
well-defined fields on $\Sigma_t$, 
\ban
H[\N] \an=\an \int d^2x \N \{
          \fr \ep^{ijk}\E^{\mu}_i\E^{\nu}_j F_{\mu\nu k}
             + \fr E\ep^{ijk}\E^{\mu}_in_j\Psi_{\mu k} 
             - mE^2\bp\psi \} \\
H_{\mu}[N^{\mu}] \an=\an \int d^2x N^{\mu} \{
               -\E^{\nu i} F_{\mu\nu i} 
               - \fr En_i\Psi^i_{\mu} \} \\
G_i[\La^i] \an=\an \int d^2x \La^i \{
              D_{\mu}\E^{\mu}_i + \fr En_i\bp\psi \}.
\ean
where $\N$, $N^{\mu}$ and $\La^i$ are smearing fields.

The constraint algebra are 
\ba
\{G_i[\La^i],G_j[\Gamma^j]\} \an=\an G_i[\ep^{ijk}\La_j\Gamma_k] \nonu
\{G_i[\La^i],H[\N]\} \an=\an 0 \nonu
\{G_i[\La^i],H_{\mu}[N^{\mu}]\} \an=\an 0 \nonu
\{H_{\mu}[N^{\mu}],H_{\nu}[M^{\nu}]\} \an=\an 
H_{\mu}[{\cal L}_{\vec{N}}M^{\mu}] + G_i[N^{\mu}M^{\nu}(F_{\mu\nu}^i
+ \fr \ep_{\mu\nu}n^i\bp\psi(n\cdot G))] \nonu
\{H_{\mu}[N^{\mu}],H[\N]\} \an=\an H[{\cal L}_{\vec{N}}\N] 
+ G_i[\N N^{\mu}\ep^{ijk}(\E^{\nu}_jF_{\mu\nu k}
+ \fr En_j\Psi_{\mu k} \nonu
\an \an \qquad\qquad\qquad + mE\ep_{\mu\nu} \E^{\nu}_jn_k\bp\psi 
- \fr \ep_{\mu\nu}\E^{\nu}_jn_k\bp\psi(n\cdot G))] \nonu
\{H[\N],H[\M]\} \an=\an 
H_{\mu}[\E^{\mu}_i\E^{\nu i}(\N\pa_{\nu}\M-\M\pa_{\nu}\N)]
\ea
Note that in these algebra, some quadratic terms of constraint $G_i$ 
appear in the right hand side, which do not appear in the case of 
the gravity including matter in 3+1 dimensions~\cite{ART}
and the non-Hermitian theory in 2+1 dimensions~\cite{KSY}. 
In classical level since these terms weakly vanish, 
the constraints are actually first class. 
So we think there is no problem.  But we do not know 
whether these terms have an effect in quantum theory.

We conclude this section with comment on the quantization of this theory. 
Since the Dirac brackets of the canonical variables are complicated, 
it is difficult to perform quantization. 
Because of the fact that $A_{\mu}^i$ does not commute with $\psi$, 
$\bp$ and itself, we can no longer represent $A_{\mu}^i$ by a multiplication 
operator, which is different from the case in~\cite{KSY,KY}. 
Vergeles~\cite{Ver} also have carried out the canonical formalism 
of the 2+1 dimensional gravity coupled with the Dirac field 
and obtained the same result that the Dirac brackets of the canonical 
variables are complicated. He has also found that 
the quadratic terms appear in the constraint algebra. 
In~\cite{Ver} the quantization have been carried out on the basis of 
the dynamic quantization method. 
To perform this, the Dirac field modes with gauge invariant creation and 
annihilation operators are selected. The gauge invariant states are built by 
using the gauge invariant fermion creation operators similarly to 
the usual construction of states in any Fock space.
In contrast with this method, 
we stand to transform the canonical variables for simplifying 
their Dirac bracket. 
This procedure is discussed in the next section.

\section{New variables}

In the last section we found that the Dirac brackets of 
the canonical variables are complicated. This complication cause 
some difficulty. For example, we can no longer construct 
quantum theory in the connection representation. 

In order to simplify the Dirac brackets of canonical variables, 
we introduce the following complex variables: 
\be
\A_{\mu}^i := A_{\mu}^i - \fr \ep_{\mu\nu}\ep^{ijk}\E^{\nu}_j\bp\ga_k\psi,
\label{new}
\ee
which commutes with itself and $\psi$. 
However, it does not commute with $\bp$, 
which then is not a suitable canonical variable. 
So we use $\pi := -E\bp\n$ as the canonical variable which replaces $\bp$. 
The non-vanishing Dirac brackets of new variables are 
\be
\{\E^{\mu}_i(x),\A_{\nu}^j(y)\} = q^{\mu}_{\nu}\eta^j_i \de^2(x,y), \ \ 
\{\pi_{\alpha}(x),\psi_{\beta}(y)\} = \de_{\alpha\beta}\de^2(x,y),
\label{Dirac}
\ee
which are the same Dirac brackets in the non-Hermitian case. 
By means of this simplification, we can avoid the difficulty indicated above. 
So we can represent $\A^i_{\mu}$ by a multiplication operator 
in quantum theory. 

Now we recast the constraints in terms of the new variables. 
The constraints (6) are rewritten as 
\ban
H \an=\an \Ha + \fr \ep_{ijk}n^i\pi\n\ga^j\psi\G^k \label{ha} \ap 0 \\
H_{\mu} \an=\an \Ha_{\mu} 
- \fr E^{-1}\ep_{\mu\nu}\ep_{ijk}\E^{\nu i}\pi\n\ga^j\psi\G^k \ap 0
\label{ve} \\
G_i \an=\an \G_i \ap 0 \label{gauss}, 
\ean
where
\ban
\Ha \an=\an \fr \ep^{ijk}\E^{\mu}_i\E^{\nu}_j \F_{\mu\nu k}
          -\E^{\mu}_i\pi\ga^i\D_{\mu}\psi 
          - mE\pi\n\psi + \frac{3}{4}(\pi\psi)^2 \label{Ha} \\
\Ha_{\mu} \an=\an -\E^{\nu}_i\F_{\mu\nu}^i + \pi \D_{\mu}\psi \\
\G_i \an=\an \D_{\mu}\E^{\mu}_i - \fr\pi\ga_i\psi,
\ean
where $\F_{\mu\nu}^i$ is the curvature tensor of $\A^i_{\mu}$ and 
$\D_{\mu}\psi$ is the covariant derivative defined by 
$\D_{\mu}\psi:=(\pa_{\mu} +\fr \A_{\mu}^i\ga_i)\psi$. 
In this calculation 
we used the Fierz transformation, $(\pi\n\psi)^2=(\pi\psi)^2$. 
In terms of the new variables, $H$ and $H_{\mu}$ become 
non-polynomial, 
and non-polynomial terms are proportional to $\G_i$.
When $\G_i \ap 0$, $\Ha$ and $\Ha_{\mu}$ are weekly equal to $H$ and $H_{\mu}$ 
respectively. 
So in this theory, 
$\Ha$ and $\Ha_{\mu}$ which are polynomial in 
terms of the new variables can be employed as the constraints: 
\be
\Ha \ap 0, \ \ \Ha_{\mu} \ap 0. \label{newconstraint}
\ee
Note that the new constraints (\ref{gauss}) and (\ref{newconstraint})
are almost the same as the 
ones of the non-Hermitian case in 2+1 dimensions~\cite{KSY} 
and Ashtekar formalism including the Dirac field in 3+1 dimensions~\cite{ART}. 
The only difference is that 
$(\pi\psi)^2$ appears in the Hamiltonian constraint $\Ha$. 
The effect of this term will be discussed in section 5. 

Using the new variables, the Hamiltonian is described as follows, 
\be
{\cal H} = -(\N \Ha + N^{\mu}\Ha_{\mu} + \Lambda^i\G_i), 
\ee
where
\be
\Lambda^i \equiv A_0^i + \fr\N \ep^{ijk}n_j\pi\n\ga_k\psi
- \fr N^{\mu}E^{-1}\ep_{\mu\nu}\ep^{ijk}\E^{\nu}_j\pi\n\ga_k\psi.
\ee
The Hamiltonian takes a form of linear combination of the constraints 
with multipliers. 
Note that, as a result of getting the polynomial constraints, 
the multiplier of the Gauss-law constraint is different from the $A_0^i$ 
of (\ref{Hamiltonian}).

Now we discuss the algebra of these constraints. 
Using the Dirac brackets (\ref{Dirac}), the constraint algebra are 
\ba
\{\G_i[\La^i],\G_j[\Gamma^j]\} \an=\an \G_i[\ep^{ijk}\La_j\Gamma_k] \nonu
\{\G_i[\La^i],\Ha[\N]\} \an=\an 0 \nonu
\{\G_i[\La^i],\Ha_{\mu}[N^{\mu}]\} \an=\an 0 \nonu
\{\Ha_{\mu}[N^{\mu}],\Ha_{\nu}[M^{\nu}]\} \an=\an 
\Ha_{\mu}[{\cal L}_{\vec{N}}M^{\mu}] 
+ \G_i[N^{\mu}M^{\nu}\F_{\mu\nu}^i] \nonu
\{\Ha_{\mu}[N^{\mu}],\Ha[\N]\} \an=\an 
\Ha[{\cal L}_{\vec{N}}\N] \nonu
\an\an + \G_i[\N N^{\mu}(\ep^{ijk}\E^{\nu}_j\F_{\mu\nu k}
- \pi \ga^i \D_{\mu}\psi + m\ep_{\mu\nu}\ep^{ijk}
\E^{\nu}_j\pi\ga_k\psi)] \nonu
\{\Ha[\N],\Ha[\M]\} \an=\an 
\Ha_{\mu}[\E^{\mu}_i\E^{\nu i}(\N\pa_{\nu}\M-\M\pa_{\nu}\N)] \label{algebra}.
\ea
Note that, in contrast with the case based on the real variables, 
no quadratic terms of $\G_i$ appear in the right hand side. 
This algebra (\ref{algebra}) is similar to the non-Hermitian case 
except for mass term, so that the structure of the constraint algebra 
is not changed by the extra term $(\pi\psi)^2$ in (\ref{Ha}). 

As a result of introducing the new canonical variables, 
we constructed a theory which has simple Dirac brackets. 
However, since the complex variables are introduced, 
the reality of the theory 
is not manifest. So we must impose reality conditions.

\section{Reality conditions}

In order to ensure that a complex theory is equivalent to 
a real one, we must impose the reality conditions 
like in the 3+1 dimensional gravity. 
If we ignore the reality conditions, that is, if we consider $\A_{\mu}^i$ 
and corresponding momentum $\E^{\mu}_i$ as complex, 
we can no longer recover the real theory. 
So we must impose the reality conditions which restrict the phase space. 

In our case, 
the canonical variables $\E^{\mu}_i$ and $\A^i_{\mu}$ are not independent 
of their complex conjugates but must satisfy 
the reality conditions 
\be
{\E^{\mu}_i}^{\da} = \E^{\mu}_i, \ \ 
{\A_{\mu}^i}^{\da} = \A_{\mu}^i + E^{-1}\ep_{\mu\nu}\ep^{ijk}
                     \E^{\nu}_j\pi\n\ga_k\psi. \label{reality1}
\ee
The latter condition of $\A^i_{\mu}$ ensure that the original variable 
$A^i_{\mu}$, which is related to $\A^i_{\mu}$ by (\ref{new}), is real. 
Furthermore, when spinor fields are included, 
additional conditions for the reality of observable currents are 
needed~\cite{ART}. 
In our case the reality conditions for spinor fields are 
given by 
\be
(\pi\E^{\mu}_i\ga^i\psi)^{\da}= - \pi\E^{\mu}_i\ga^i\psi, \ \ 
(\pi\E^{\mu}_i\ga^i\pi)^{\da}=E^2(\psi\E^{\mu}_i\ga^i\psi).\label{reality2}
\ee
Using these conditions, the constraints $\Ha$,$\Ha_{\mu}$ and $\G_i$ 
obey the following relations 
\ba
\Ha^{\da} \an=\an \Ha + \ep^{ijk}n_i\pi\n\ga_j\psi \G_k \nonu
\Ha_{\mu}^{\da} \an=\an \Ha_{\mu} + E^{-1}\ep_{\mu\nu}\ep^{ijk}
                        \E^{\nu}_i\pi\n\ga_j\psi \G_k \nonu
\G_i^{\da} \an=\an \G_i. \label{G}
\ea
Note that the constraints $\Ha$ and $\Ha_{\mu}$ are complex. 
Therefore the Hamiltonian takes a form of linear combination 
of the complex constraints. 
In order to satisfy the reality conditions at any time, 
the Hamiltonian must be real. So the Lagrange multipliers 
in the Hamiltonian are not totally arbitrary but must satisfy 
\ba
\N^{\da} \an=\an \N, \ \ {N^{\mu}}^{\da}=N^{\mu} \nonu
{\La^i}^{\da} \an=\an \La^i + \N \ep^{ijk}n_j\pi\n\ga_k\psi
-N^{\mu}E^{-1}\ep_{\mu\nu}\ep^{ijk}\E^{\nu}_j\pi\n\ga_k\psi. 
\label{reality3}
\ea
If the reality conditions (\ref{reality1}) and (\ref{reality2}) 
are satisfied at initial time and 
the multipliers obey the relations (\ref{reality3}), 
the equivalence between the real theory 
and the complex one is guaranteed. 
Note that the reality conditions (\ref{reality1}) and 
the multiplier conditions (\ref{reality3}) are similar to the ones 
in the case of pure gravity in 3+1 dimensions~\cite{KF}. 
In both theories, complex connection is introduced to simplify the 
canonical structure. 
In our case the aim is to simplify the Dirac brackets 
of canonical variables. 
In the case of pure gravity in 3+1 dimensions, on the other hand, 
the complex connection is used 
to get the constraints in polynomial form. 

Now we consider reality conditions which are different from (19) 
in appearance.
First we take $\La^i$ as an arbitrary complex function in the Hamiltonian. 
Therefore the Hamiltonian is in general complex. 
As new reality conditions, we impose that the triad $\E^{\mu}_i$ is real, 
\be
\im \E^{\mu}_i = 0 \label{Ereal1}
\ee
and that the time derivative of $\E^{\mu}_i$ is also real, 
\be
\im \dot{\E}{}^{\mu}_i=0. \label{Ereal2}
\ee
As regards spinor fields, the spinor reality conditions 
(\ref{reality2}) are required. 
Using the Gauss-law constraint, the spinor reality conditions 
(\ref{reality2}) and the triad reality condition (\ref{Ereal1}), 
it follows that (\ref{Ereal2}) reduces to 
\be
\im \A_{\mu}^i = -\fr E^{-1}\ep_{\mu\nu}\ep^{ijk}\E^{\nu}_j\pi\n\ga_k\psi, 
\label{A}
\ee
which is equivalent to the second condition in (\ref{reality1}). 
On the other hand, it is possible to solve (\ref{Ereal2}) with regard to 
$\La^i$, 
\be
\im \La^i = \fr \N \ep^{ijk}n_j\pi\n\ga_k\psi
+ N^{\mu} \im \A_{\mu}^i \label{lambda}. 
\ee
This condition restricts a part of gauge freedom, and this situation 
is similarly to the 3+1 dimensional gravity~\cite{YS}. 
Using (\ref{A}), we see that the condition (\ref{lambda}) is the same as 
the last condition in (\ref{reality3}). 
Thus the reality conditions (\ref{Ereal1}) and (\ref{Ereal2}) 
are equivalent to the those of (\ref{reality1}) and (\ref{reality2}). 

We count the number of (real) degrees of freedom 
assuming there is only one Dirac field. 
The complex canonical variables $\A_{\mu}^i$ and $\E^{\mu}_i$ have 
12 independent components respectively. 
When we consider that $\psi$ and $\pi$ are independent, 
there are 8 independent components of spinor fields. 
The constraints are also complex in general. But due to the reality conditions 
(\ref{reality1}) and (\ref{reality2}), all the constraints 
are not independent of each other. 
>From (\ref{G}) we see that $\G_i$ is real and that the imaginary parts 
of $\Ha$ and $\Ha_{\mu}$ are proportional to $\G_i$. 
Thus 6 independent constraints remain. 
With 12 reality conditions (\ref{reality1}) 
and 4 for the spinor field (\ref{reality2}), we find that the remaining 
number of degrees of freedom is $(24+8)-2\times 6 -(12+4) = 4$. 
Because there is no graviton in 2+1 dimensions, 
this corresponds to the degrees of freedom of spinor field $\psi$. 
So it follows that the reality conditions (\ref{reality1}) and 
(\ref{reality2}) are suitable and reproduce the correct 
number of degrees of freedom.

\section{Conclusion}

We studied (2+1)-dimensional gravity coupled with the Dirac field. 
In contrast with the non-Hermitian case~\cite{KSY}, 
the Dirac brackets of the canonical variables are complicated. 
We found that quadratic terms of the constraint $G_i$ appear 
in the constraint algebra. No such term appears also in the case of 
the gravity including matter in 3+1 dimensions~\cite{ART}
and the non-Hermitian theory in 2+1 dimensions~\cite{KSY}. 
In order to simplify the Dirac brackets of the canonical variables, 
we introduced the complex variables, and found that quadratic terms of 
the Gauss-law constraint disappear 
and that the constraint algebra becomes similar to 
the case of the gravity including matter in 3+1 dimensions~\cite{ART}
and the non-Hermitian theory in 2+1 dimensions~\cite{KSY}. 
But being different from both cases, the Hamiltonian constraint 
of the present theory contains 
$(\pi\psi)^2$. 

Next we considered the reality conditions. By virtue of these conditions, 
the phase space is restricted, and the original real theory is 
recovered. In order to retain the reality conditions at any time, 
it is important that the Hamiltonian is real; accordingly, the Lagrange 
multipliers are not totally arbitrary but are related to their complex 
conjugates. 
We also considered the different reality conditions, which are 
imposed only on the triad. 
We showed that these new conditions are reduced 
to the original ones after all. 
We also showed that, owing to the reality conditions, the correct 
number of degrees of freedom remains. 

Finally 
we give a short comment on the quantum theory. 
We can define the canonical operators $\A^i_{\mu}$ and $\psi$ as 
multiplicative operators, and the corresponding momentum operators 
$\E^{\mu}_i$ and $\pi$ as the derivative operators. 
We take the ordering in which 
momentum operators are placed to the right. 
In the non-Hermitian case~\cite{KSY}, 
as a solution including spinor fields for the Hamiltonian constraint, 
a trivial one $\psi^A\psi_A$ has been found. 
This solution corresponds 
to an eigenstate of the fermion number operator $\int d^2x \psi_A\pi^A$. 
In our Hermitian case, however, since the Hamiltonian constraint contains 
$(\psi\pi)^2$, $\psi^A\psi_A$ is no longer a solution. 
We are now looking for other solutions.


\noindent{\large{\bf Acknowledgments}} \\
We would like to thank Professor T Shirafuji for reading the manuscript 
and useful comments. 
We are grateful to Professor K Kamimura for valuable discussions. 
We would also like to thank members of the theory group at Saitama University 
for useful comments. 



\end{document}